\documentclass[a4paper, aps, prb, reprint, superscriptaddress, showkeys, amsmath, amssymb, nofootinbib, twoside, longbibliography]{revtex4-2}

\usepackage{dcolumn}
\usepackage{ulem}
\usepackage{bm}
\usepackage{color}
\usepackage{graphicx} 
\usepackage{multirow}
\usepackage{braket}
\usepackage{hyperref} 
\usepackage{natbib}

\newcommand{\Eq}[1]{Eq.~(\ref{#1})}
\newcommand{\Sec}[1]{Sec.~\ref{#1}}
\newcommand{\Fig}[1]{Fig.~\ref{#1}}
\newcommand{\App}[1]{Appendix~\ref{#1}}

\begin{document}

\title{Nonlinear frequency-asymmetric optical response in chiral systems}
\author{Shuhei Kanda}
\author{Satoru Hayami}
\affiliation{Graduate School of Science, Hokkaido University, Sapporo 060-0810, Japan}
\begin{abstract}
We report our theoretical results on the emergence of a nonlinear frequency-asymmetric optical response characteristic of chiral crystal systems with neither spatial inversion symmetry nor mirror symmetry.
Based on the group theoretical analysis, we show that the chirality-related second-order nonlinear optical response occurs for two different input frequencies when the low-energy model Hamiltonian includes a time-reversal-even pseudoscalar quantity, i.e., the electric toroidal monopole. 
We demonstrate its emergence by investigating a fundamental microscopic model with the chiral-type antisymmetric spin--orbit interaction on a simple cubic lattice. 
By analyzing the behavior of nonlinear optical conductivity based on the Kubo formula, we find that the response is largely enhanced when one of the frequencies is set to zero and the other is set to a resonant frequency. 
We also discuss the relaxation time dependence of the response.
\end{abstract}
\maketitle

\section{introduction} \label{sec:introduction}
Physical responses in materials are determined by their symmetries.
For example, noncentrosymmetric systems without spatial inversion symmetry show fascinating physical responses against external fields and currents, such as the piezoelectric effect~\cite{curie1880,fukada1957piezoelectric} and the Edelstein effect~\cite{edelstein1990spin, Fujimoto_PhysRevB.72.024515, yoda2018orbital}. 
Among them, the nonlinear transport against the square of the input electric field has been extensively studied in recent years~\cite{tokura2018nonreciprocal, gao2019semiclassical}, 
which includes the second harmonic generation~\cite{PhysRevLett.108.077403, wu2017giant, Gao2023QMHall}, 
nonreciprocal transport~\cite{ideue2017bulk, Rikken_PhysRevLett.87.236602, PhysRevB.105.155157, Hayami_PhysRevB.106.014420}, 
photogalvanic effect~\cite{PhysRevB.52.14636, PhysRevB.102.035440, PhysRevX.11.011001, ma2022unveiling, Huaung2021Photogal}, 
and nonlinear Hall effect~\cite{PhysRevLett.115.216806, Nandy_PhysRevB.100.195117, ma2022unveiling,PhysRevLett.127.277201,Liu_PhysRevLett.127.277202, doi:10.7566/JPSJ.91.014701,PhysRevB.107.155109,li2023current}.
The microscopic understanding of not only exploring various mechanisms but also enhancing such responses has been an active research area in condensed matter physics. 

The multipole description provides a way to link macroscopic phenomena with microscopic electronic degrees of freedom from the symmetry~\cite{PhysRevB.98.165110,PhysRevB.98.245129,PhysRevResearch.2.043081,kusunose2022generalization,hayami2024unified}.  
There are four types of multipoles with different spatial inversion ($\mathcal{P}$) and time-reversal ($\mathcal{T}$) parities: 
electric multipole corresponding to a $\mathcal{T}$-even polar tensor, magnetic multipole corresponding to a $\mathcal{T}$-odd axial tensor, 
magnetic toroidal multipole corresponding to a $\mathcal{T}$-odd polar tensor, 
and electric toroidal multipole corresponding to a $\mathcal{T}$-even axial tensor.
Among them, the odd-rank electric and magnetic toroidal multipoles and even-rank magnetic and electric toroidal multipoles show an odd-parity transformation in terms of the $\mathcal{P}$ operation, which are referred to as the odd-parity multipoles. 
In the case of the second-order nonlinear conductivity characterized by a rank-3 polar tensor, the rank 0--3 odd-parity multipoles play an important role~\cite{PhysRevB.98.165110,PhysRevB.104.054412}; when
the expectation values of the multipole moments are finite, the corresponding tensor component can become nonzero.
In addition, the multipole description also provides the essential model parameters in the conductive phenomena, as demonstrated in the case of a monolayer of the ferroelectric material SnTe to have the electric dipole degree of freedom~\cite{doi:10.7566/JPSJ.91.014701}.
Moreover, the close relationship between the nonlinear physical phenomena and multipoles has been clarified, such as the nonreciprocal transport under the magnetic toroidal dipole~\cite{PhysRevB.105.155157, Hayami_PhysRevB.106.024405}, 
intrinsic nonlinear Hall effect under the magnetic quadrupole~\cite{PhysRevB.107.155109}, and nonlinear transverse magnetization under the electric toroidal dipole~\cite{doi:10.7566/JPSJ.92.043701}. 

In the present study, we propose yet another interesting situation that is induced by the rank-0 electric toroidal monopole under chiral crystal structures with neither spatial inversion symmetry nor mirror symmetry~\cite{PhysRevB.98.165110, PhysRevLett.129.116401, hayami2024analysis}.
Although fascinating transport phenomena have been unveiled in chiral crystals like the phonon magnetic chiral effect and macroscopic spin response~\cite{Rikken_PhysRevLett.87.236602, nomura2019phonon, Inui_PhysRevLett.124.166602, nakazawa2024nonlinear, PhysRevLett.132.056302,ishito2023truly}, 
we here focus on second-order nonlinear optical conductivities with different input frequencies.
First, we investigate the nonlinear optical response characteristic of chiral crystals based on symmetry and multipole analyses.
We show that the electric toroidal monopole degree of freedom activated under chiral crystals is symmetry-related to frequency-asymmetric optical responses against the second-order electric field. 
Next, we analyze such an optical response for a fundamental model with antisymmetric spin--orbit interaction (ASOI) that originates from the chiral structures. 
By examining the nonlinear Kubo formula, we derive analytical expressions for the second-order nonlinear conductivities, where the ASOI plays an essential role.
Furthermore, we discuss the behavior against the input frequencies, and as a result,
we find that the optical response is largely enhanced when one of the frequencies is the DC limit, while the other corresponds to a resonant frequency.
We also investigate the dependence of the relaxation time.

The organization of this paper is as follows:
In \Sec{sec:Second-order optical conductivity tensor}, we present the relationship between the second-order nonlinear optical conductivity tensor and multipoles based on the group theory.
We show that the electric toroidal monopole induced under chiral structures contributes to the nonlinear optical conductivity when the input frequencies are different. 
In \Sec{sec:Model and method}, we examine the behavior of the nonlinear frequency-asymmetric optical conductivity by analyzing the fundamental tight-binding model incorporating the chiral-type ASOI. 
Starting from the nonlinear Kubo formula, we derive the analytical expressions of the optical conductivity. 
In \Sec{sec:Numerical Results}, we present the numerical results with an emphasis on the dependence of input frequencies and relaxation times.
Finally, in \Sec{sec:Summary}, we summarize the paper.

\section{Second-order optical conductivity tensor} \label{sec:Second-order optical conductivity tensor}
The second-order optical conductivity tensor represents the current response proportional to the square of the input electric field. That is expressed by
\begin{equation}
    \begin{aligned}
        j^{\mu}(\omega)=\sum_{\alpha_1,\alpha_2} \int\frac{d\omega_1}{2\pi} \int\frac{d\omega_2}{2\pi} 2\pi\delta(\omega-\omega_1-\omega_2)\\
        \times \sigma^{\mu;\alpha_1\alpha_2}(\omega_1, \omega_2)E^{\alpha_1}(\omega_1)E^{\alpha_2}(\omega_2) ,
    \end{aligned}
\end{equation}
where $E^{\alpha_1}(\omega_1)$, $j^{\mu}(\omega)$, and $\sigma^{\mu;\alpha_1\alpha_2}(\omega_1,\omega_2)$ stand for the $\alpha_1=x,y,z$ component of the electric field with the frequency $\omega_1$, 
the $\mu=x,y,z$ component of the electric current with the frequency $\omega=\omega_1+\omega_2$, and second-order nonlinear conductivity tensor characterized by the third-rank polar tensor, respectively.
In the following, we introduce the frequency-asymmetric component by decomposing 
$\sigma^{\mu;\alpha_1\alpha_2}(\omega_1,\omega_2)$ 
into symmetric and antisymmetric components in \Sec{subsec:Frequency-asymmetric} and discuss the relevance to the multipoles in \Sec{subsec:Correspondence to multipole} based on the group theory.

\subsection{Frequency-asymmetric nonlinear conductivity} \label{subsec:Frequency-asymmetric}
The second-order nonlinear optical conductivity tensor is symmetric with respect to the simultaneous interchange of the components of the electric field and the input frequencies as follows:
\begin{align}
    \sigma^{\mu;\alpha_1\alpha_2}(\omega_1,\omega_2)=\sigma^{\mu;\alpha_2\alpha_1}(\omega_2,\omega_1).
\end{align}
Meanwhile, it can be decomposed into a symmetric part ($\sigma_{\mathrm{S}}$) and an antisymmetric part ($\sigma_{\mathrm{A}}$) for the interchange of either the components of the electric field or the input frequencies~\cite{PhysRevX.11.011001}, which are expressed as
\begin{align}
    \sigma^{\mu;\alpha_1\alpha_2}(\omega_1,\omega_2)
    =\sigma^{\mu;\alpha_1\alpha_2}_{\mathrm{S}}(\omega_1,\omega_2)+\sigma^{\mu;\alpha_1\alpha_2}_{\mathrm{A}}(\omega_1,\omega_2),
\end{align}
with
\begin{equation}
    \begin{alignedat}{4}
        \sigma^{\mu;\alpha_1\alpha_2}_{\mathrm{S}}(\omega_1,\omega_2) &= &&\sigma^{\mu;\alpha_1\alpha_2}_{\mathrm{S}}(\omega_2,\omega_1)&&= &&\sigma^{\mu;\alpha_2\alpha_1}_{\mathrm{S}}(\omega_1,\omega_2),  \\
        \label{eq:freintasym}
        \sigma^{\mu;\alpha_1\alpha_2}_{\mathrm{A}}(\omega_1,\omega_2) &=-&&\sigma^{\mu;\alpha_1\alpha_2}_{\mathrm{A}}(\omega_2,\omega_1)&&=-&&\sigma^{\mu;\alpha_2\alpha_1}_{\mathrm{A}}(\omega_1,\omega_2).
    \end{alignedat}
\end{equation}

In the situation where two input frequencies are symmetric, such as the DC limit of the electric conductivity with $\omega_1=\omega_2=0$~\cite{doi:10.7566/JPSJ.91.014701,RevModPhys.82.1539,PhysRevB.107.155109,PhysRevLett.112.166601} and the second harmonic generation with $\omega_1=\omega_2\ne0$~\cite{wu2017giant,PhysRevB.97.235446,PhysRevLett.108.077403,Gao2023QMHall}, 
only $\sigma^{\mu;\alpha_1\alpha_2}_{\mathrm{S}}(\omega_1,\omega_2)$ contributes to the conductivity; $\sigma^{\mu;\alpha_1\alpha_2}_{\mathrm{A}}(\omega_1,\omega_2)$ identically vanishes from the symmetry.
Meanwhile, in the situation where the two input frequencies are different, i.e., $\omega_1\ne\omega_2$, as represented by the photogalvanic effect ($\omega_1=-\omega_2$)~\cite{ma2022unveiling,ma2022unveiling,PhysRevX.11.011001,PhysRevB.102.035440}
or the AC voltage with DC bias ($\omega_1=0,\omega_2\ne0$)~\cite{minussi2021dc},
$\sigma^{\mu;\alpha_1\alpha_2}_{\mathrm{A}}(\omega_1,\omega_2)$ can become nonzero in addition to $\sigma^{\mu;\alpha_1\alpha_2}_{\mathrm{S}}(\omega_1,\omega_2)$.

\subsection{Correspondence to multipole} \label{subsec:Correspondence to multipole}
From the symmetry viewpoint, the relevant multipoles to induce $\sigma^{\mu;\alpha_1\alpha_2}(\omega_1,\omega_2)$ are understood by the addition rule of the input electric field and the output electric current~\cite{PhysRevB.98.165110,PhysRevB.104.054412,PhysRevB.92.155138}.
Owing to the property in \Eq{eq:freintasym}, the corresponding electric fields for $\sigma^{\mu;\alpha_1\alpha_2}_{\mathrm{S}}(\omega_1,\omega_2)$ are given by the rank-$0$ form with $\bm{E}(\omega_1)\cdot \bm{E}(\omega_2)$ and the rank-2 form with $E^{\alpha_1}(\omega_1)E^{\alpha_2}(\omega_2)+E^{\alpha_2}(\omega_1)E^{\alpha_1}(\omega_2)$, 
whereas that for $\sigma^{\mu;\alpha_1\alpha_2}_{\mathrm{A}}(\omega_1,\omega_2)$ is given by the rank-1 form with $\bm{E}(\omega_1) \times \bm{E}(\omega_2)$. 
Since the direct product of the rank-$N$ electric field and rank-$1$ electric current leads to odd-parity multipoles from rank $|N-1|$ to $N+1$,
the relevant multipoles for $\sigma^{\mu;\alpha_1\alpha_2}_{\mathrm{S}}(\omega_1,\omega_2)$ are the rank-1 to rank-3 odd-parity multipoles, while those for $\sigma^{\mu;\alpha_1\alpha_2}_{\mathrm{A}}(\omega_1,\omega_2)$ are the rank-0 to rank-2 odd-parity multipoles. 
Thus, the rank-3 odd-parity multipoles only contribute to $\sigma^{\mu;\alpha_1\alpha_2}_{\mathrm{S}}(\omega_1,\omega_2)$, while the rank-0 odd-parity multipoles only contribute to $\sigma^{\mu;\alpha_1\alpha_2}_{\mathrm{A}}(\omega_1,\omega_2)$. 
In the following, we focus on the relationship between $\sigma^{\mu;\alpha_1\alpha_2}_{\mathrm{A}}(\omega_1,\omega_2)$ and multipoles; the relationship between $\sigma^{\mu;\alpha_1\alpha_2}_{\mathrm{S}}(\omega_1,\omega_2)$ and multipoles has been discussed in Ref.~\cite{PhysRevB.104.054412}.

The relationship between the tensor components and multipoles for $\sigma^{\mu;\alpha_1\alpha_2}_{\mathrm{A}}(\omega_1,\omega_2)$ is given by 
\begin{align}
\label{eq:sigmaA}
    \sigma_{\mathrm{A}}=
    \begin{pmatrix}
        { Y_{0} - Y_{u} + Y_{v}} & { X_{z} + Y_{xy}}        & {-X_{y} + Y_{zx}} \\
        {-X_{z} + Y_{xy}}        & { Y_{0} - Y_{u} - Y_{v}} & { X_{x} + Y_{yz}} \\
        { X_{y} + Y_{zx}}        & {-X_{x} + Y_{yz}}        & { Y_{0} + 2Y_{u}}
     \end{pmatrix},
\end{align}
where the matrix is represented by $ (j^x,j^y,j^z)^{\mathrm{T}} =  \sigma _{\mathrm{A}} (E^yE^z,E^zE^x,E^xE^y)^{\mathrm{T}} $. 
$Y_0$, $(X_x, X_y, X_z)$, and $(Y_{u}, Y_{v}, Y_{yz}, Y_{zx}, Y_{xy})$ ($u=3z^2-r^2$ and $v=x^2-y^2$) denote the odd-parity monopole, dipole, and quadrupole, respectively, where $X$ and $Y$ represent the polar and axial quantities, respectively. 
The specific correspondence between multipoles and tensor components is given by
\begin{subequations}
    \begin{align}
        Y_0           &=                    \chi^{\mathrm{M}(1\times1)}, \label{eq:defY0}                                                              \\
        (X_x,X_y,X_z) &=             \left( \chi^{\mathrm{D}(1\times1)}_{x} ,\chi^{\mathrm{D}(1\times1)}_{y},\chi^{\mathrm{D}(1\times1)}_{z} \right)  ,\\
        Y_u           &=                   3\chi^{\mathrm{Q}(1\times1)}_{zz} - \sum_{i} \chi^{\mathrm{Q}(1\times1)}_{ii}                              ,\\
        Y_v           &= \frac{1}{2} \left( \chi^{\mathrm{Q}(1\times1)}_{xx} - \chi^{\mathrm{Q}(1\times1)}_{yy} \right)                               ,\\
        (Y_{yz},Y_{zx},Y_{xy}) &=    \left( \chi^{\mathrm{Q}(1\times1)}_{yz},\chi^{\mathrm{Q}(1\times1)}_{zx},\chi^{\mathrm{Q}(1\times1)}_{xy} \right),
    \end{align}
\end{subequations}
where
\begin{subequations}
    \begin{align}
        \chi^{\mathrm{M}(1\times1)}      &= \frac{1}{6} \sum_{ijk} {\epsilon^{ijk} \sigma^{i;jk}}, \\
        \chi^{\mathrm{D}(1\times1)}_{i}  &= \frac{1}{4} \sum_{j}   (\sigma^{j;ij}-\sigma^{j;ji}),  \\
        \chi^{\mathrm{Q}(1\times1)}_{ij} &= \frac{1}{4} \sum_{kl}  (\epsilon^{ikl}\sigma^{j;kl}+\epsilon^{jkl}\sigma^{i;kl}), 
    \end{align}
\end{subequations}
where $\epsilon^{ijk}$ is the Levi-Civita tensor. 
Thus, when the system accompanies the rank-0 electric toroidal (magnetic) monopole in the presence of the $\mathcal{T}$ ($\mathcal{PT}$) symmetry, $\sigma^{\mu;\alpha_1\alpha_2}_{\mathrm{A}}(\omega_1,\omega_2)$ can be finite.
Meanwhile, $\sigma^{\mu;\alpha_1\alpha_2}_{\mathrm{s}}(\omega_1,\omega_2)$ is not induced under such rank-0 monopoles.
Since the electric toroidal monopole is regarded as a microscopic order parameter in chiral structures belonging to the Sohncke groups~\cite{PhysRevB.98.165110, PhysRevLett.129.116401}, the pure frequency-asymmetric response $\sigma^{\mu;\alpha_1\alpha_2}_{\mathrm{A}}(\omega_1,\omega_2)$ is expected in chiral materials.

Let us consider a situation where $\sigma^{\mu;\alpha_1\alpha_2}_{\mathrm{A}}(\omega_1,\omega_2)$ can become nonzero by taking a cubic system under the point group $O_{\mathrm{h}}$ as an example. 
We suppose that the time-reversal symmetry is preserved; $X$ and $Y$ in \Eq{eq:sigmaA} correspond to the electric multipole and electric toroidal multipole, respectively.
In this situation, since the electric field $\bm{E}$ belongs to the irreducible representation $T_{1\mathrm{u}}$, the irreducible representation of the second order of $\bm{E}$ is given by 
\begin{align}
    T_{1\mathrm{u}} \otimes T_{1\mathrm{u}} = A_{1\mathrm{g}} \oplus E_{\mathrm{g}} \oplus T_{1\mathrm{g}} \oplus T_{2\mathrm{g}}.
\end{align}
Among them, $A_{1\mathrm{g}}$ corresponds to $\bm{E}(\omega_1)\cdot \bm{E}(\omega_2)$, 
$E_{\mathrm{g}} \oplus T_{2\mathrm{g}}$ to $E^{\alpha_1}(\omega_1)E^{\alpha_2}(\omega_2)+E^{\alpha_2}(\omega_1)E^{\alpha_1}(\omega_2)$, and $T_{1\mathrm{g}}$ to $\bm{E}(\omega_1) \times \bm{E}(\omega_2)$. 
Furthermore, since the electric current also belongs to the irreducible representation $T_{1\mathrm{u}}$, the irreducible representation of $\sigma^{\mu;\alpha_1\alpha_2}_{\mathrm{A}}(\omega_1,\omega_2)$ is given by 
\begin{align}
    T_{1\mathrm{g}} \otimes T_{1\mathrm{u}}=A_{1\mathrm{u}} \oplus E_{\mathrm{u}} \oplus T_{1\mathrm{u}} \oplus T_{2\mathrm{u}}.
\end{align}
There are nine independent degrees of freedom: $A_{1\mathrm{u}}$ corresponds to the electric toroidal monopole $G_0$, $T_{1\mathrm{u}}$ corresponds to the electric dipole $(Q_x, Q_y, Q_z)$, 
and $E_{\mathrm{u}} \oplus T_{2\mathrm{u}}$ corresponds to the electric toroidal quadrupole $(G_{u}, G_{v}, G_{yz}, G_{zx}, G_{xy})$.
Thus, when the $O_{\rm h}$ symmetry lowers to the $O$ symmetry so that $A_{1\mathrm{u}}$ belongs to the identity irreducible representation, $\sigma^{\mu;\alpha_1\alpha_2}_{\mathrm{A}}(\omega_1,\omega_2)$ can be induced. 
Especially, since the electric dipoles, the electric toroidal quadrupoles, and the electric octupoles do not belong to the identity irreducible representation in this case, only the component 
$\sigma^{x;yz}_{\mathrm{A}}(\omega_1,\omega_2)=\sigma^{y;zx}_{\mathrm{A}}(\omega_1,\omega_2)=\sigma^{z;xy}_{\mathrm{A}}(\omega_1,\omega_2)=-\sigma^{x;zy}_{\mathrm{A}}(\omega_1,\omega_2)=-\sigma^{y;xz}_{\mathrm{A}}(\omega_1,\omega_2)=-\sigma^{z;yx}_{\mathrm{A}}(\omega_1,\omega_2)$ 
can become nonzero; other components in $\sigma^{\mu;\alpha_1\alpha_2}_{\mathrm{A}}(\omega_1,\omega_2)$ and $\sigma^{\mu;\alpha_1\alpha_2}_{\mathrm{S}}(\omega_1,\omega_2)$ are identically zero.

\section{Model and method} \label{sec:Model and method}
As described in the previous section, the chiral system under the $O$ symmetry exhibits the pure nonlinear frequency-asymmetric optical conductivity $\sigma^{\mu;\alpha_1\alpha_2}_{\mathrm{A}}(\omega_1,\omega_2)$.
In order to examine the behavior of $\sigma^{\mu;\alpha_1\alpha_2}_{\mathrm{A}}(\omega_1,\omega_2)$, we construct a fundamental tight-binding model in \Sec{subsec:Model Hamiltonian}. 
Then, we show the analytical expressions of $\sigma^{\mu;\alpha_1\alpha_2}_{\mathrm{A}}(\omega_1,\omega_2)$ based on the Kubo formula in \Sec{subsec:Kubo Formula}.

\subsection{Model Hamiltonian}\label{subsec:Model Hamiltonian}
We consider the tight-binding Hamiltonian on a three-dimensional simple cubic lattice under the point group $O$, which is given by
\begin{align}
\label{eq:Hamtotal}
    \mathcal{H}=\mathcal{H}_{\mathrm{hop}}+\mathcal{H}_{\mathrm{ASOI}}.
\end{align}
The first term $\mathcal{H}_{\mathrm{hop}}$ represents the hopping Hamiltonian, which is expressed as
\begin{equation}
    \begin{aligned}
        \mathcal{H}_{\mathrm{hop}} &= \sum_{\bm{k}}\sum_{s}   \varepsilon_0(\bm{k})c^{\dagger}_{\bm{k}s}c_{\bm{k}s} ,\\
        \varepsilon_0(\bm{k}) &= -2t(\cos{k_x}+\cos{k_y}+\cos{k_z}),
    \end{aligned}
\end{equation}
where $c^{\dagger}_{\bm{k}s}$ and $c_{\bm{k}s}$ are the fermionic creation and annihilation operaters 
of the wave number $\bm{k}$ and the spin $s=\uparrow,\downarrow$.
$\varepsilon_0(\bm{k})$ represents the nearest-neighbor hopping with the magnitude of $t$ on the cubic lattice;  the lattice constant is set to be $a=1$ for simplicity.

The second term in \Eq{eq:Hamtotal} represents the ASOI Hamiltonian, which is given by
\begin{equation}
    \begin{aligned}
        \mathcal{H}_{\mathrm{ASOI}} &= \alpha\sum_{\bm{k}}\sum_{s,s'}
        \bm{g}(\bm{k})\cdot{\bm{\sigma}_{ss'}}c^{\dagger}_{\bm{k}s}c_{\bm{k}s'},\\
        \bm{g}(\bm{k}) &= (\sin{k_x},\sin{k_y},\sin{k_z}) ,
    \end{aligned}
\end{equation}
where $\bm{\sigma}=(\sigma_x,\sigma_y,\sigma_z)$ is the vector of Pauli matrices in spin space.
$\bm{g}(\bm{k})$ is characterized by the antisymmetric function with respect to the wave vector $\bm{k}$, which is the so-called $g$ vector~\cite{PAFrigeri2006}.
Owing to the chiral point group $O$, the wave vector $\bm{k}$ is parallel to the spin $\bm{\sigma}$, i.e., $\bm{k}\parallel \bm{\sigma}$, which has the same symmetry as the electric toroidal monopole $G_0$.
Thus, the system exhibits the antisymmetric spin-split band structure~\cite{PhysRevB.98.165110}.
The microscopic origin of the ASOI is the interplay between the relativistic spin--orbit coupling and parity mixing between orbitals.

\subsection{Kubo Formula}\label{subsec:Kubo Formula}
The nonlinear frequency-asymmetric optical conductivity is calculated by using the Kubo formula.
As shown in \App{sec:Decomposition}, $\sigma^{\mu;\alpha_1\alpha_2}(\omega_1,\omega_2)$ can be generally decomposed into intraband (i) and interband (e) contributions under the length gauge as follows
~\cite{PhysRevX.11.011001,PhysRevB.99.045121,PhysRevB.61.5337,PhysRevB.52.14636}:
\begin{align}
  \sigma^{\mu;\alpha_1\alpha_2}=\sigma^{\mu;\alpha_1\alpha_2}_{\mathrm{ii}}
                               +\sigma^{\mu;\alpha_1\alpha_2}_{\mathrm{ei}}
                               +\sigma^{\mu;\alpha_1\alpha_2}_{\mathrm{ie}}
                               +\sigma^{\mu;\alpha_1\alpha_2}_{\mathrm{ee}},
\end{align}
where we omit the frequency dependence for simplicity.
Among the terms, the second term ${\sigma_\mathrm{ei}}$ is related to the Berry curvature dipole term~\cite{PhysRevLett.115.216806}.
As discussed below, ${\sigma_\mathrm{ei}}$ appears under the ETM order; we focus on the role of ${\sigma_\mathrm{ei}}$ in the following analysis.
The expression of ${\sigma_\mathrm{ei}}$ is given by~\cite{PhysRevX.11.011001,PhysRevB.99.045121} 
\begin{equation}
    \begin{aligned}
        &\sigma^{\mu;\alpha_1\alpha_2}_{\mathrm{ei}}(\omega_1,\omega_2)\\
        &\quad=-\frac{e^3}{\hbar^2V}\sum_{\bm{k}}\sum_{a,b}^{\varepsilon_a\ne\varepsilon_b}\frac{p_a v^{\alpha_2}_{aa}}{\omega_2+i\gamma}
        \left( \varepsilon_{ab}d^{+}_{ab}\mathcal{G}^{\alpha_1\mu}_{ab}-\frac{i}{2}\varepsilon_{ab}d^{-}_{ab}\mathcal{F}^{\alpha_1\mu}_{ab} \right)\\
        &\qquad+((\omega_1,\alpha_1) \leftrightarrow (\omega_2,\alpha_2)),
    \end{aligned}
\end{equation}
where the wave vector and frequency dependence is omitted [e.g., $\varepsilon_a=\varepsilon_a(\bm{k})$].
$e$, $\hbar$, and $V$ are the elementary charge, the reduced Planck constant, and the system volume, respectively.
$\gamma=1/\tau$ is the scattering rate, characterized as the inverse of the relaxation time $\tau$ under the relaxation time approximation.
$\varepsilon_a(\bm{k})$ is a band energy with wave vector $\bm{k}$ and band index $a$, and $\varepsilon_{ab}=\varepsilon_a-\varepsilon_b$ is a energy difference.
$p_a=df(\varepsilon_a)/d\varepsilon$ is the derivative of the Fermi distribution function 
\begin{align}
    f_a=f(\varepsilon_a)=\frac{1}{e^{(\varepsilon_a-\mu)/k_\mathrm{B}T}+1},
\end{align}
where $\mu$ is the chemical potential.
$v^{\alpha}_{ab}$ is a Bloch representation for the velocity operator defined by
\begin{align}
    v^{\alpha}_{ab} = \Braket{u_a(\bm{k})| \frac{1}{\hbar} \frac{\partial H_{\bm{k}}}{\partial k_{\alpha}} |u_b(\bm{k})},
\end{align}
where $H_{\bm{k}}$ and $\ket{u_a(\bm{k})}$ are $\bm{k}$-resolved Hamiltonian and the eigenstate, respectively.
$\mathcal{G}^{\alpha_1\alpha_2}_{ab}$ and $\mathcal{F}^{\alpha_1\alpha_2}_{ab}$ are referred to as the band-resolved quantum metric tensor and Berry curvature tensor, respectively~\cite{doi:10.7566/JPSJ.91.014701,cheng2010quantum,PhysRevX.11.011001}, 
whose expressions are given by
\begin{equation}
    \begin{aligned}
        \mathcal{G}^{\alpha_1\alpha_2}_{ab}
        =\frac{\hbar^2}{2}
        \frac{v^{\alpha_1}_{ab}v^{\alpha_2}_{ba}+v^{\alpha_2}_{ab}v^{\alpha_1}_{ba}}{(\varepsilon_a-\varepsilon_b)^2},\\
        \mathcal{F}^{\alpha_1\alpha_2}_{ab}
        =i\hbar^2
        \frac{v^{\alpha_1}_{ab}v^{\alpha_2}_{ba}-v^{\alpha_2}_{ab}v^{\alpha_1}_{ba}}{(\varepsilon_a-\varepsilon_b)^2}.
    \end{aligned}
\end{equation}
$d^{\pm}_{ab}=d^{\pm}_{ab}(\omega_1+\omega_2,2\gamma)$ is the (anti-)symmetrized resonant Green's function
~\cite{PhysRevX.11.011001,PhysRevB.52.14636,PhysRevB.97.235446,PhysRevB.96.035431}: 
\begin{align}\label{eq:green}
    d^{\pm}_{ab}(\omega,\gamma)
    =\frac{1}{2}\left( \frac{1}{\hbar\omega-\varepsilon_{ab}+i\hbar\gamma}\pm\frac{1}{\hbar\omega+\varepsilon_{ab}+i\hbar\gamma} \right).
\end{align}

By calculating the $Y_0$ component of ${\sigma_\mathrm{ei}}$ in \Eq{eq:defY0}, we obtain 
\begin{equation}\label{eq:defY0Kubo}
    \begin{aligned}
        {Y_0}^{(\mathrm{ei})}(\omega_1,\omega_2)
        =i \frac{e^3}{6\hbar^2 V} \left(\frac{1}{\omega_2+i\gamma}-\frac{1}{\omega_1+i\gamma} \right) \\
        \times\sum_{\bm{k}}\sum_{a,b}^{\varepsilon_a\ne\varepsilon_b}(p_a \bm{v}_{aa}) \cdot \left( \varepsilon_{ab} d^{-}_{ab} \bm{\mathcal{B}}_{ab} \right),
    \end{aligned}
\end{equation}
where $\bm{\mathcal{B}}_{ab}$ is the vector representation of the band-resolved Berry curvature:
\begin{align}
    \mathcal{B}^{\mu}_{ab}=\sum_{\alpha_1\alpha_2}\epsilon^{\mu\alpha_1\alpha_2}\mathcal{F}^{\alpha_1\alpha_2}_{ab},
   \quad\bm{\mathcal{B}}_{ab}=2i\hbar^2\frac{\bm{v}_{ab}\times\bm{v}_{ba}}{(\varepsilon_a-\varepsilon_b)^2}.
\end{align}
It is noted that $\bm{\mathcal{B}}_{ab}(\bm{k})$ vanishes in the presence of the $\mathcal{PT}$ symmetry owing to the relation $\bm{\mathcal{B}}_{ab}(\bm{k}) = -\bm{\mathcal{B}}_{ab}(\bm{k})$~\cite{PhysRevX.11.011001}. 
This indicates that magnetic monopole does not contribute to $Y_0^{(\mathrm{ei})}$ in the model in \Eq{eq:Hamtotal}; only the electric toroidal monopole contributes.

In the present model in \Eq{eq:Hamtotal}, only the tensor component 
$\sigma^{x;yz}_{\mathrm{A}}(\omega_1,\omega_2)=\sigma^{y;zx}_{\mathrm{A}}(\omega_1,\omega_2)=\sigma^{z;xy}_{\mathrm{A}}(\omega_1,\omega_2)=-\sigma^{x;zy}_{\mathrm{A}}(\omega_1,\omega_2)=-\sigma^{y;xz}_{\mathrm{A}}(\omega_1,\omega_2)=-\sigma^{z;yx}_{\mathrm{A}}(\omega_1,\omega_2)$ 
becomes nonzero. 
The specific expressions of $\sigma^{z;xy}(\omega_1,\omega_2)$ are analytically calculated by 
\begin{equation}\label{eq:calculated}
  \begin{aligned}
      \sigma^{z;xy}_{\mathrm{ei}}(\omega_1,\omega_2)&={Y_0}^{(\mathrm{ei})}(\omega_1,\omega_2)\\
      &=i \frac{e^3}{3\hbar^2 V} \left( \frac{1}{\omega_2+i\gamma}-\frac{1}{\omega_1+i\gamma} \right) \\
      &\qquad \times\sum_{\bm{k}}^{\varepsilon_{+} \ne \varepsilon_{-}} \frac{f_{+}-f_{-}}{\varepsilon_{+}-\varepsilon_{-}} J_{\alpha\bm{g}} \left(2 d_{+-}^{+} \right)^2,
  \end{aligned}
\end{equation}
where $\varepsilon_{\pm}=\varepsilon_0\pm \alpha g\,(g=\sqrt{\bm{g}\cdot\bm{g}})$ is the eigenenergy with the band index $\pm$ of the $\bm{k}$-resolved expression in \Eq{eq:Hamtotal} and $J_{\alpha\bm{g}}=\alpha^3 \cos k_x \cos k_y \cos k_z $; 
the detailed derivation is presented in \App{sec:Derivation of the analytical expression}. 
This expression clearly indicates the importance of the ASOI in inducing $\sigma^{z;xy}_{\mathrm{ei}}(\omega_1,\omega_2)$; $\sigma^{z;xy}_{\mathrm{ei}}(\omega_1,\omega_2)=0$ for $\alpha=0$. 
Hereafter, we omit the subscript A in the conductivity, since there is no contribution from the symmetric component.

\section{Numerical Results} \label{sec:Numerical Results}
\begin{figure*}[t]
  \centering
  \includegraphics[width=1.0\linewidth]{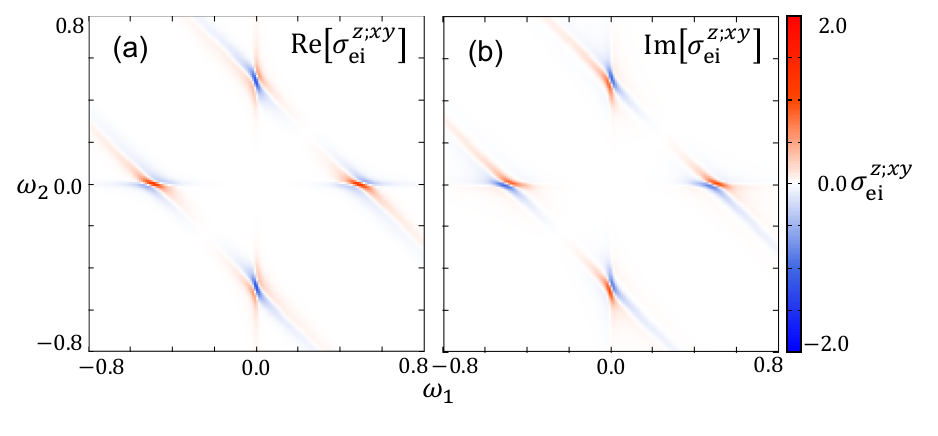}
  \caption{\label{fig:frequency map}
  The color map showing the behavior of the real part (a) and imaginary part (b) of the second-order optical response tensor $\sigma^{z;xy}_{\mathrm{ei}}(\omega_1,\omega_2)$ with respect to the input frequencies $(\omega_1, \omega_2)$.
  We set $\gamma=0.01$ and $N=128^3$.
  The real (imaginary) part of $\sigma^{z;xy}_{\mathrm{ei}}(\omega_1,\omega_2)$ exhibits symmetric (antisymmetric) behavior in terms of the line $\omega_2=\omega_1$ and antisymmetric (symmetric) behavior in terms of $\omega_2=-\omega_1$. 
  It is noted that there is no response for second harmonic generation ($\omega_1=\omega_2$) and the photogalvanic effect ($\omega_1=-\omega_2$).
  }
\end{figure*}

In the following numerical calculations, we set $t=1$ and $\alpha=0.2$. 
We also set $e=\hbar=k_{\mathrm{B}}=1$, the temperature $T=0.01$, and the chemical potential $\mu=-4$. 
The number of the $\bm{k}$ mesh $N=V/a^3$ is appropriately taken so that the results reach convergence.

\subsection{Frequency dependence}\label{subsec:Frequency dependence}
First, we investigate the dependence of $\sigma^{z;xy}_{\mathrm{ei}}(\omega_1,\omega_2)$ on the two input frequencies $\omega_1$ and $\omega_2$ by numerically evaluating the expression in \Eq{eq:calculated}. 
The scattering rate is set as $\gamma=0.01$, and the $\bm{k}$ mesh is taken as $N=128^3$. 
Figure~\ref{fig:frequency map} shows a color map of the real and imaginary parts of $\sigma^{z;xy}_{\mathrm{ei}}(\omega_1,\omega_2)$ in the plane of $(\omega_1,\omega_2)$. 
$\sigma^{z;xy}_{\mathrm{ei}}(\omega_1,\omega_2)$ exhibits a characteristic behavior along the lines $\omega_2=\pm\omega_1$:
\begin{equation}
  \begin{aligned}\label{eq:fredepreg}
      \sigma^{z;xy}_{\mathrm{ei}}(\omega_2,\omega_1)&=-\sigma^{z;xy}_{\mathrm{ei}}(\omega_1,\omega_2),\\
      \sigma^{z;xy}_{\mathrm{ei}}(-\omega_2,-\omega_1)&=-\left(\sigma^{z;xy}_{\mathrm{ei}}(\omega_1,\omega_2)\right)^*.
  \end{aligned}
\end{equation}
This relation indicates that $\sigma^{z;xy}_{\mathrm{ei}}(\omega_1,\omega_2)$ vanishes for $\omega_2=\pm \omega_1$, i.e., $\sigma^{z;xy}_{\mathrm{ei}}(\omega, \omega)=0$ and $\sigma^{z;xy}_{\mathrm{ei}}(\omega,-\omega)=0$; 
there is no second harmonic generation and optical photogalvanic effect.
This relation arises from the two factors. 
One is the antisymmetric nature of the tensor $\sigma_{\mathrm{A}}$ for the interchange of the frequency.
The other is the relation satisfying, $\sigma^{z;xy}_{\mathrm{ei}}(-\omega_1, -\omega_2) = \left(\sigma^{z;xy}_{\mathrm{ei}}(\omega_1, \omega_2)\right)^*$ due to $\bm{E}(-\omega) = \bm{E}^*(\omega)$ and $\bm{j}(-\omega) = \bm{j}^*(\omega)$.
The absence of $\sigma^{z;xy}_{\mathrm{ei}}(\omega, -\omega)=0$ is also understood from the analytical expression.
By supposing the clean limit, i.e., $\gamma \approx 0$, \Eq{eq:defY0Kubo} is rewritten as 
\begin{align}
  Y_0^{(\mathrm{ei})}(\omega, -\omega) \propto i \sum_{\bm{k}} \sum_{a,b}^{\varepsilon_a \ne \varepsilon_b} f_a \frac{\partial}{\partial \bm{k}} \cdot \bm{\mathcal{B}}_{ab} + O(\gamma^2),
\end{align}
where $\sum_{b}^{\varepsilon_a \ne \varepsilon_b} {\partial_{\bm{k}}} \cdot \bm{\mathcal{B}}_{ab}$ represents the divergence of the Berry curvature. 
Since such a singularity appears in pairs of opposite signs in the three-dimensional Brillouin zone, $\sigma^{z;xy}_{\mathrm{ei}}$ vanishes for bulk~\cite{NIELSEN1983389,Vanderbilt_2018}. 
Thus, $\sigma^{z;xy}_{\mathrm{ei}}(\omega, -\omega)$ vanishes in the present model, although it is symmetry-allowed.

\begin{figure}[b]
  \begin{center}
      \includegraphics[width=1.0\linewidth]{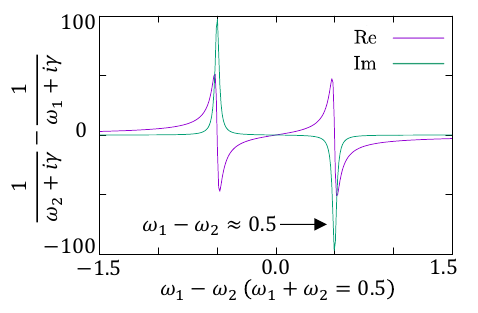}
      \caption{\label{fig:frequency difference}
      The dependence of the factor in the \Eq{eq:frequency difference} on the frequency difference $\omega_1-\omega_2$ for $\gamma=0.01$ and $\omega_1+\omega_2=0.5$. 
      The purple line represents the real part (Re), while the green line represents the imaginary part (Im). 
      The imaginary part shows the peaks when $\omega_1 - \omega_2 \approx \pm(\omega_1 + \omega_2)$. 
      At these frequencies, the real part becomes zero but peaks nearby.
      }
  \end{center}
\end{figure}

\begin{figure*}[t]
  \begin{center}
      \includegraphics[width=1.0\linewidth]{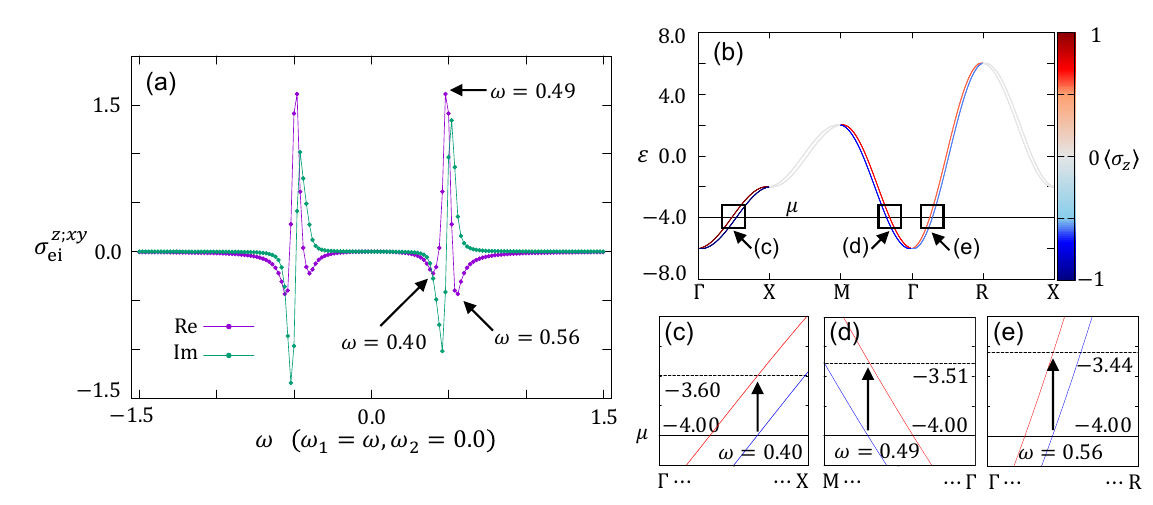}
      \caption{\label{fig:frequency sum}
      (a) $\omega$ dependence of $\sigma^{z;xy}_{\mathrm{ei}}(\omega,0)$ in \Eq{eq:calculated}, where we set $\gamma=0.01$ and $N=128^3$.
      The purple line represents the real part (Re), while the green line represents the imaginary part (Im).
      In the real part, peaks are observed at $\omega = \pm 0.40, \pm 0.49, \pm 0.56$.
      (b)
      The energy dispersion of $H_{\bm{k}}$ along the high-symmetry lines: $\Gamma$ means $\bm{k} = (0,0,0)$, M means $\bm{k} = (0,\pi,\pi)$, 
      R means $\bm{k} = (\pi,\pi,\pi)$, and 
      X means $\bm{k} = (0,0,\pi)$.
      The color bar represents the expectation value of the $z$ spin component with $\bm{k}$ and the band index $\pm$, and the black solid horizontal line indicates the chemical potential $\mu = -4$. 
      (c)--(e) The enlarged data for the three areas enclosed by squares in (b) are shown in (c), (d), and (e). 
      The transition energies from the chemical potential correspond to the peaks in (a).
      }
  \end{center}
\end{figure*}

In both Figs.~\ref{fig:frequency map}(a) and (b), the responses are enhanced along lines with $\omega_1+\omega_2 \simeq 0.5$.
Furthermore, within these lines, the magnitude of the responses is maximized in the vicinity when one frequency is DC, i.e., $\omega_1=0$ or $\omega_2=0$.
This behavior is understood by evaluating the dependencies on the sum ($\omega_1+\omega_2$) and the difference ($\omega_1-\omega_2$) of the two frequencies. 
From \Eq{eq:calculated}, the factor with the frequency dependence is represented by
\begin{align}\label{eq:frequency difference}
  \frac{1}{\omega_2+i\gamma}-\frac{1}{\omega_1+i\gamma}=\frac{\omega_1-\omega_2}{(\omega_1\omega_2-\gamma^2)+i\gamma(\omega_1+\omega_2)}.
\end{align}
We show the $\omega_1 - \omega_2$ dependence of the term in \Eq{eq:frequency difference} at $\omega_1 + \omega_2 = 0.5$ in \Fig{fig:frequency difference}.
The results show that the term in \Eq{eq:frequency difference} exhibits the peaks around the frequencies satisfying the condition $\omega_1 - \omega_2 = \pm (\omega_1 + \omega_2)$ for small $\gamma$. 
Since this condition is satisfied for either $\omega_1 = 0$ or $\omega_2 = 0$, $\sigma^{z;xy}_{\mathrm{ei}}$ is also enhanced for either $\omega_1 = 0$ or $\omega_2 = 0$.
Thus, a large response is expected when the input electric field corresponds to the AC voltage with DC bias satisfying 
\begin{align}
  \omega_1=\omega\,(\gg\gamma),\quad \omega_2=0.
\end{align}
In such a situation, the term in \Eq{eq:frequency difference} is simplified as
\begin{align}
  \frac{1}{\omega_2+i\gamma}-\frac{1}{\omega_1+i\gamma}=-i\frac{1}{\gamma}+O(\gamma)\approx -i\tau.
\end{align}
Thus, $\sigma^{z;xy}_{\rm ei}$ is proportional to $\tau$.
A similar phenomenon, the nonlinear Hall effect induced by a DC, has also been reported~\cite{li2023current}.

To further examine the behavior of $\sigma^{z;xy}_{\mathrm{ei}}(\omega,0)$, we plot its $\omega$ dependence in \Fig{fig:frequency sum}(a).
Particularly focusing on the real part, peaks are observed at specific frequencies $\omega = \pm0.40, \pm0.49, \pm0.56$. 
These peaks are caused by the resonance with the band structure.
Indeed, $d^{-}_{ab}$ in \Eq{eq:green} is transformed in the clean limit $(\gamma\rightarrow+0)$ as
\begin{equation}
  \begin{aligned}\label{resonance}
      \mathrm{Im}[d^{-}_{ab}(\omega,\gamma)]\approx -\pi[\delta(\hbar\omega-\varepsilon_{ab})-\delta(\hbar\omega+\varepsilon_{ab})],
  \end{aligned}
\end{equation}
which indicates that the important contribution occurs when the frequency corresponds to the energy difference between the two bands, i.e., 
$\varepsilon_{+} - \varepsilon_{-} \simeq \hbar\omega $.
Such a correspondence is found in the band structures, as shown in \Fig{fig:frequency sum}(b);
the color bar represents the expectation value of the $z$ spin component $\braket{u_{\pm}(\bm{k})|\sigma_z|u_{\pm}(\bm{k})}$, 
and the black solid horizontal line indicates the chemical potential $\mu = -4$. 
We also show the enlarged figures for the energy dispersions crossing the Fermi level in the high-symmetry lines in Figs.~\ref{fig:frequency sum}(c)--\ref{fig:frequency sum}(e).
The frequencies to enhance $\sigma^{z;xy}_{\mathrm{ei}}(\omega,0)$, which correspond to $\omega=\pm 0.40$, $\pm 0.49$, and $\pm 0.56$, are explained by the energy differences split by the ASOI.
In other words, the resonance between the spin-split bands by the ASOI plays an important role in obtaining large frequency-asymmetric responses in chiral systems.

\subsection{Relaxation time dependence}
\begin{figure*}[t]
  \centering
  \includegraphics[width=0.9\linewidth]{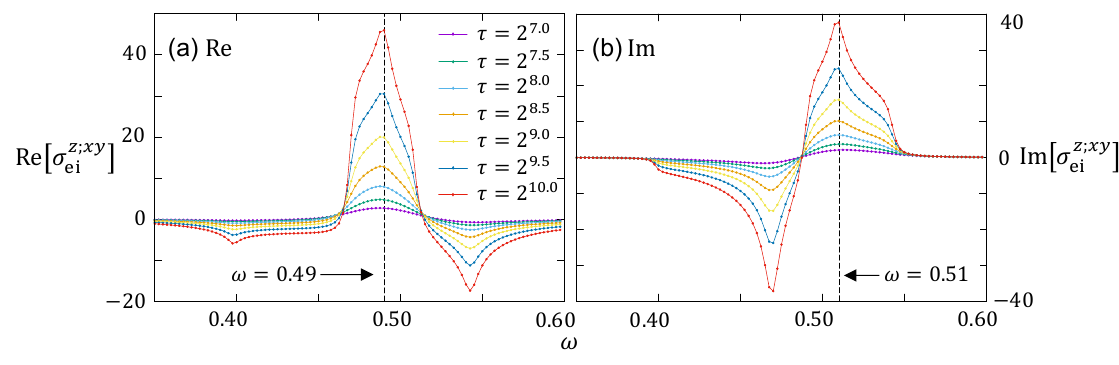}
  \caption{\label{fig:frequency relax}
  $\omega$ dependence of (a) the real and (b) imaginary parts of $\sigma_{\rm ei}^{z;xy}$ in \Eq{eq:calculated} for the relaxation times $\tau = 2^{7}, 2^{7.5}, \cdots, 2^{10}$ combined with \Eq{eq:frequency relax} at $N=2048^3$. 
  In the real part, the dominant peak is observed at $\omega=0.49$, while in the imaginary part, one of the peaks is observed at $\omega=0.51$.
  As the relaxation time increases, the response becomes larger in both cases.
  }
\end{figure*}
Next, we investigate the relaxation-time dependence of $\sigma^{z;xy}_{\mathrm{ei}}(\omega_1,\omega_2)$ in \Eq{eq:calculated}.
Similarly to the previous section, we focus on the situation where $\sigma^{z;xy}_{\mathrm{ei}}(\omega_1,\omega_2)$ is enhanced by setting $\omega_1 \simeq \omega$ and $\omega_2 \simeq 0$.
In order to analytically deal with the relaxation time dependence, we parameterize $\omega_1$ and $\omega_2$ as follows: 
\begin{equation}
  \begin{aligned}\label{eq:frequency relax}
      \omega_1 &= \frac{\omega}{2} \left( 1+\sqrt{1- \left(\frac{2}{\tau\omega} \right) ^2} \right),\\
      \omega_2 &= \frac{\omega}{2} \left( 1-\sqrt{1- \left(\frac{2}{\tau\omega} \right) ^2} \right),
  \end{aligned}
\end{equation}
where $\omega_1 \simeq \omega$ and $\omega_2 \simeq 0$ for $\omega\tau \gg 1$. 
It is noted that $\omega_1+ \omega_2 = \omega$.
\Fig{fig:frequency relax} shows the $\omega$ dependence of $\sigma^{z;xy}_{\mathrm{ei}}(\omega_1,\omega_2)$ in \Eq{eq:calculated} for various relaxation times as $\tau = 2^{7}, 2^{7.5}, \ldots, 2^{10}$ with the system size $N=2048^3$.
As the relaxation time increases, both real and imaginary components become larger, as shown in Figs.~\ref{fig:frequency relax}(a) and \ref{fig:frequency relax}(b), respectively.

\begin{figure*}[t]
  \centering
  \includegraphics[width=0.9\linewidth]{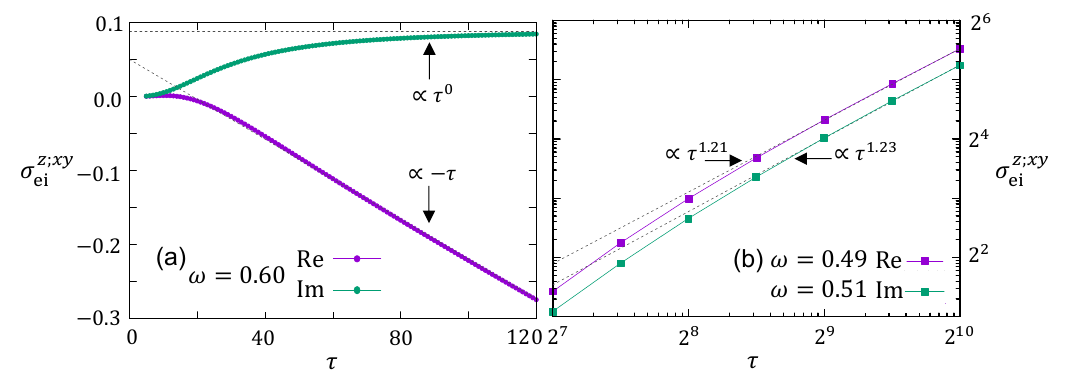}
  \caption{\label{fig:relaxtime}
  Relaxation time dependence for (a) the non-resonant frequency with $\omega=0.60$ and (b) the resonant frequency at $N=128^3$.
  In (b), the frequency is set as $\omega=0.49$ for the real part and $\omega=0.51$ for the imaginary part.
  The purple and green lines represent the behavior of the real and imaginary parts, respectively.
  In (a), the real part of $\sigma_{\rm ei}^{z;xy}$ shows $\propto -\tau$, while the imaginary part shows $(\propto 1)$ for large $\tau$. 
  In (b), both the real and imaginary parts show non-trivial relaxation time dependence: ${\rm Re}[\sigma_{\rm ei}^{z;xy}]\propto\tau^{1.21}$ and ${\rm Im}[\sigma_{\rm ei}^{z;xy}]\propto\tau^{1.23}$.
  The dotted lines are the fitting lines for large $\tau$. 
  }
\end{figure*}

In order to examine the relaxation time dependence, we consider the two situations.
First, we consider the situation when the non-resonant frequency is applied. 
The result for $N = 128^3$ and $\omega = 0.60$ is shown in \Fig{fig:relaxtime}(a), which indicates that the $\tau$ dependence of $\sigma^{z;xy}_{\mathrm{ei}}$ gradually approaches the following values: 
\begin{align}
  \mathrm{Re}[\sigma^{z;xy}_{\mathrm{ei}}] \propto -\tau ,\quad \mathrm{Im}[\sigma^{z;xy}_{\mathrm{ei}}] \propto 1,
\end{align}
in the clean limit $\tau \rightarrow \infty$ ($\gamma \rightarrow 0$).
This behavior is understood from the $\tau$ dependence of the factor $d^{-}_{ab}$ included in $\sigma^{z;xy}_{\mathrm{ei}}$, which is expressed as
\begin{equation}
  \begin{aligned}
      \mathrm{Re}[d^{-}_{ab}(\omega,\gamma) ] &\propto \,     1        + O(\gamma^2),\\
      \mathrm{Im}[d^{-}_{ab}(\omega,\gamma) ] &\propto -\frac{1}{\tau} + O(\gamma^3).
  \end{aligned}
\end{equation}

In contrast to the non-resonant frequency, the $\tau$ dependence is complicated in the situation where the resonant frequency is applied.
For the peak positions of the real-part conductivity at $\omega = 0.49$ and the imaginary-part conductivity at $\omega = 0.51$, we investigate each $\tau$ dependence, as shown in \Fig{fig:relaxtime}(b). 

We numerically obtain the non-trivial $\tau$ dependence as follows:
\begin{align}
  \mathrm{Re}[\sigma^{z;xy}_{\mathrm{ei}}] \propto \tau^{1.21} ,\quad \mathrm{Im}[\sigma^{z;xy}_{\mathrm{ei}}] \propto \tau^{1.23}.
\end{align}
Although it is difficult to obtain such exponents in an analytical way, this complicated behavior might be attributed to the following reasons.
The $\tau$ dependence directly arises from the factor of the frequency difference as in  \Eq{eq:frequency difference} ($\propto -i\tau$) and $d^{-}_{ab}$. 
The term $d^{-}_{ab}$ exhibits a delta function-like contribution $\tau/\hbar$ from the range $|\varepsilon_{+} - \varepsilon_{-} - \hbar\omega| < \hbar/\tau$. 
This means that as $\tau$ increases, the peak increases proportionally to $\tau$, while the peak range narrows. 
The reduction in the number of states contributing to the peak due to the decreasing peak range results in a non-trivial $\tau$ dependence characterized by $p_a$.
As a result, non-trivial relaxation time dependence appears in the resonant-frequency case.

\section{Summary} \label{sec:Summary}
In summary, we have proposed the possibility of nonlinear frequency-asymmetric optical response phenomena under chiral crystal structures. 
By performing the group theoretical analysis, we have shown the relationship between the second-order nonlinear frequency-asymmetric optical conductivity and odd-parity multipole, where the electric toroidal monopole contributes to the conductivity in chiral systems.
Then, we investigated the behavior of the second-order nonlinear frequency-asymmetric optical conductivity by the microscopic model calculations.
By analyzing the Kubo formula for the fundamental tight-binding model with the chiral-type ASOI, we discussed the frequency and relaxation time dependence.  
With respect to the frequency dependence, we have shown that the conductivity is largely enhanced when one of the frequencies is set to zero and the other is set to a resonant frequency. 
With respect to the relaxation-time dependence, we have shown its difference whether the frequency is resonant or not. 

The present second-order nonlinear frequency-asymmetric optical conductivity is expected to appear in noncentrosymmetric materials, such as the chiral systems with the rank-0 electric toroidal monopole in 
tellurium~\cite{vorob1979optical, shalygin2012current, yoda2015current, furukawa2017observation}, 
CrNb$_3$S$_6$~\cite{miyadai1983magnetic, Inui_PhysRevLett.124.166602}, 
and chiral molecules~\cite{inda2024quantification}, 
the polar systems with the rank-1 electric dipole in SnTe~\cite{chang2016discovery, Xu_PhysRevB.95.235434, kim2019prediction}, 
and gyrotropic systems with the rank-2 electric toroidal quadrupole in Cd$_2$Re$_2$O$_7$~\cite{yamaura2002low, hiroi2018pyrochlore, Hayami_PhysRevLett.122.147602, hirai2022successive}. 
Among them, the cubic materials under the point group $O$ are promising to investigate the pure frequency-asymmetric responses since only the electric toroidal monopole is activated in such a case. 
The candidate materials are Co$_8$Zn$_9$Mn$_3$~\cite{yu2018transformation}, SrSi$_2$~\cite{singh2018tunable}, Mg$_3$Ru$_2$~\cite{doi:10.1021/ic800387a}, Mn$_3$Cu$_3$O$_8$~\cite{https://doi.org/10.1002/advs.202303694}.

\begin{acknowledgments}
This research was supported by 
JSPS KAKENHI Grants Numbers JP21H01037, JP22H00101, JP22H01183, JP23H04869, JP23K03288, JP23K20827, 
and by JST CREST (JPMJCR23O4).  
\end{acknowledgments}

\appendix
\section{Derivation of the analytical expression} \label{sec:Derivation of the analytical expression}
We show the derivation of the analytical expression in \Eq{eq:calculated} in the main text.
We start from the tight-binding Hamiltonian in \Eq{eq:Hamtotal}, which is rewritten in the general form as
\begin{equation}
  \begin{aligned}
    \mathcal{H} &= \sum_{\bm{k}}\sum_{s,s'}[H_{\bm{k}}]_{ss'}c^{\dagger}_{\bm{k}s}c_{\bm{k}s'},\\
    H_{\bm{k}}  &= R_0(\bm{k})+\bm{R}(\bm{k})\cdot\bm{\sigma}.
  \end{aligned}
\end{equation}
\Eq{eq:Hamtotal} is obtained by setting $R_0(\bm{k})=\varepsilon_{0}(\bm{k})$ and $\bm{R}(\bm{k})=\alpha\bm{g}(\bm{k})$.
By diagonalizing $H_{\bm{k}}$, the eigenenergy $\varepsilon_{\pm}(\bm{k})$ is given as follow:
\begin{equation}
  \begin{aligned}
    \varepsilon_{\pm}(\bm{k}) &= R_0(\bm{k})\pm R(\bm{k})\quad(R=\sqrt{\bm{R \cdot R}}),\\
  \end{aligned}
\end{equation}
The eigenstates with the wave vector $\bm{k}$ and the band index $\pm$ are
\begin{equation}
  \begin{aligned}
    \ket{u_{+}(\bm{k})}=
    \begin{pmatrix}     
      { e^{ -i\frac{\phi_{\bm{R}}}{2}} \cos{\frac{\varphi_{\bm{R}}}{2}}}  \\
      { e^{  i\frac{\phi_{\bm{R}}}{2}} \sin{\frac{\varphi_{\bm{R}}}{2}}}   
    \end{pmatrix},
    \ket{u_{-}(\bm{k})}=
    \begin{pmatrix}     
      { e^{ -i\frac{\phi_{\bm{R}}}{2}} \sin{\frac{\varphi_{\bm{R}}}{2}}}  \\
      {-e^{  i\frac{\phi_{\bm{R}}}{2}} \cos{\frac{\varphi_{\bm{R}}}{2}}}   
    \end{pmatrix},
  \end{aligned}
\end{equation}
where the vector representation is expressed in spin space.
$\varphi_{\bm{R}}$ and $\phi_{\bm{R}}$ are defined by
\begin{equation}
  \begin{aligned}
    \bm{R}=R(
    \cos{\phi_{\bm{R}}}\sin{\varphi_{\bm{R}}},\,
    \sin{\phi_{\bm{R}}}\sin{\varphi_{\bm{R}}},\,
    \cos{\varphi_{\bm{R}}}).
  \end{aligned}
\end{equation}
The matrix elements of the Pauli matrix defined in the above basis are
\begin{equation}
  \begin{aligned}
    \sigma^{\alpha}_{\pm\pm}(\bm{k})=\pm\frac{R_{\alpha}}{R},\,
    \sigma^{\alpha}_{\pm\mp}(\bm{k})=-\frac{R_{z}}{R}\mathrm{e}^{\alpha}_{r}+\frac{r}{R}\mathrm{e}^{\alpha}_{z}\pm i\mathrm{e}^{\alpha}_{\phi},
  \end{aligned}
\end{equation}
where
\begin{equation}
  \begin{gathered}
    r=R\sin{\varphi_{\bm{R}}},\\
    \bm{\mathrm{e}}_{r} =\cos{\phi_{\bm{R}}}\bm{\mathrm{e}}_{x}+\sin{\phi_{\bm{R}}}\bm{\mathrm{e}}_{y},\quad
    \bm{\mathrm{e}}_{\phi}=-\sin{\phi_{\bm{R}}}\bm{\mathrm{e}}_{x}+\cos{\phi_{\bm{R}}}\bm{\mathrm{e}}_{y}.
  \end{gathered}
\end{equation}
Then, the velocity operator is described by $\hbar v^\alpha=\partial_{\bm{k}}R_0+(\partial_{\bm{k}}R_{\alpha})\sigma^{\alpha}$, 
whose matrix elements are given by
\begin{equation}
  \begin{aligned}
    &v^{\alpha}_{\pm\pm}(\bm{k})=\frac{1}{\hbar}\left(\frac{\partial R_0}{\partial k_\alpha}\pm\frac{\partial R}{\partial k_\alpha}\right),\\
    &v^{\alpha}_{\pm\mp}(\bm{k}) \\
    &\,=\frac{1}{\hbar}\left(\frac{\partial R}{\partial r}\frac{\partial R_z}{\partial k_\alpha}-\frac{\partial R}{\partial R_z}\frac{\partial r}{\partial k_\alpha}\right)
    \pm \frac{i}{\hbar}\left(\frac{\partial r}{\partial R_x}\frac{\partial R_y}{\partial k_\alpha}-\frac{\partial r}{\partial R_y}\frac{\partial R_x}{\partial k_\alpha}\right),
  \end{aligned}
\end{equation}
where the relation
\begin{align}
  dr=\frac{R_x}{r}dR_x+\frac{R_y}{r}dR_y,
\end{align}
was used. Therefore,
\begin{equation}
  \begin{aligned}
    &v^{\alpha_1}_{+-}(\bm{k})v^{\alpha_2}_{-+}(\bm{k})\\
    &\,=\frac{1}{\hbar^2}\left(\frac{\partial \bm{R}}{\partial k_{\alpha_1}}\cdot\frac{\partial \bm{R}}{\partial k_{\alpha_2}}-\frac{\partial R}{\partial k_{\alpha_1}}\frac{\partial R}{\partial k_{\alpha_2}}\right)
        -\frac{i}{\hbar^2}\frac{\bm{R}}{R} \cdot \left(\frac{\partial \bm{R}}{\partial k_{\alpha_1}}\times\frac{\partial \bm{R}}{\partial k_{\alpha_2}}\right).
  \end{aligned}
\end{equation}
In addition, the band-resolved quantum metric tensor and Berry curvature tensor are given by 
\begin{equation}
  \begin{aligned}
    \mathcal{G}^{\alpha_1\alpha_2}_{+-}(\bm{k})&=\frac{1}{8}\frac{1}{R^2}            \left(\frac{\partial \bm{R}}{\partial k_{\alpha_1}}\cdot\frac{\partial \bm{R}}{\partial k_{\alpha_2}}-\frac{\partial R}{\partial k_{\alpha_1}}\frac{\partial R}{\partial k_{\alpha_2}}\right),\\
    \mathcal{F}^{\alpha_1\alpha_2}_{+-}(\bm{k})&=\frac{1}{4}\frac{\bm{R}}{R^3} \cdot \left(\frac{\partial \bm{R}}{\partial k_{\alpha_1}}\times\frac{\partial \bm{R}}{\partial k_{\alpha_2}}\right).
  \end{aligned}
\end{equation}

Using the above results, $Y_0^{(\mathrm{ei})}$ in \Eq{eq:defY0Kubo} can be rewritten as
\begin{equation}
  \begin{aligned}
      &{Y_0}^{(\mathrm{ei})}(\omega_1,\omega_2)\\
      &\,=i \frac{e^3}{3\hbar^2 V} \left(\frac{1}{\omega_2+i\gamma}-\frac{1}{\omega_1+i\gamma} \right)\\
      &\quad\times\sum_{\bm{k}}^{\varepsilon_+\ne\varepsilon_-}\frac{f_+ - f_-}{\varepsilon_{+}-\varepsilon_{-}} (2d^+_{+-})^2 2R^2 \frac{\partial R}{\partial\bm{k}} \cdot \bm{\mathcal{B}}_{+-} ,
  \end{aligned}
\end{equation}
where integration by parts, the energy difference $\varepsilon_{+}-\varepsilon_{-}=2R$, and the relation that $\partial_{\bm{k}}\cdot\bm{\mathcal{B}}_{+-}=0$ in a two-level system were used. 
Furthermore, the relation
\begin{equation}
  \begin{aligned}
    \frac{\partial}{\partial\varepsilon_{ab}}\left(\varepsilon_{ab} d^{-}_{ab}(\omega,\gamma)\right)
    &=-\frac{\varepsilon_{ab}}{2}(2d^+_{ab}(\omega,\gamma))^2 \\
  \end{aligned}
\end{equation}
was used.

Considering $\bm{R}(\bm{k})=\alpha\bm{g}(\bm{k})=\alpha(\sin{k_x},\sin{k_y},\sin{k_z})$, we obtain
\begin{equation}
  \begin{aligned}
    \bm{\mathcal{B}}_{+-} &= \frac{1}{2g^3}(\sin{k_x}\cos{k_y}\cos{k_z},(\mathrm{cyclic.})),\\
    \frac{\partial R}{\partial\bm{k}} &= \frac{\alpha}{g}(\sin{k_x}\cos{k_x},(\mathrm{cyclic.})),\\
    \therefore 
    2R^2\frac{\partial R}{\partial\bm{k}} \cdot \bm{\mathcal{B}}_{+-} &= \alpha^3\cos{k_x}\cos{k_y}\cos{k_z}\\
                                                                      &=  J_{\alpha\bm{g}}.
  \end{aligned}
\end{equation}
As a result, we derive
\begin{equation}
  \begin{aligned}
      {Y_0}^{(\mathrm{ei})}(\omega_1,\omega_2)&=i \frac{e^3}{3\hbar^2 V} \left( \frac{1}{\omega_2+i\gamma}-\frac{1}{\omega_1+i\gamma} \right) \\
      &\quad \times\sum_{\bm{k}}^{\varepsilon_{+} \ne \varepsilon_{-}} \frac{f_{+}-f_{-}}{\varepsilon_{+}-\varepsilon_{-}} J_{\alpha\bm{g}} \left(2 d_{+-}^{+} \right)^2,
  \end{aligned}
\end{equation}
which correspond to \Eq{eq:calculated} in the main text.

\section{Decomposition into intraband and interband contributions} \label{sec:Decomposition}
We decompose the second-order nonlinear conductivity into the intraband (i) and interband (e) contributions.
The Bloch representation of the position operator is given by 
\begin{equation}
  \begin{aligned}
      \bm{r}&=\bm{r}_{\mathrm{i}}+\bm{r}_{\mathrm{e}}\\
      \bm{r}_{\mathrm{i}}&=\sum_{\bm{k}}\sum_{a  } c^{\dagger}_{\bm{k}a}i\frac{\partial c_{\bm{k}a}}{\partial\bm{k}},\quad
      \bm{r}_{\mathrm{e}} =\sum_{\bm{k}}\sum_{a,b} \bm{\xi}_{ab} c^{\dagger}_{\bm{k}a}c_{\bm{k}b},
  \end{aligned}
\end{equation}
where 
\begin{align}
  \bm{\xi}_{ab}=i\Braket{u_a(\bm{k})| \frac{\partial}{\partial \bm{k}} |u_b(\bm{k})}
\end{align}
is the Berry connection~\cite{PhysRevB.99.045121,PhysRevX.11.011001}.
$\bm{r}_{\mathrm{i}}$ represents the intraband contribution of the position operator, while $\bm{r}_{\mathrm{e}}$ represents the interband contribution.
In the length gauge, the second-order nonlinear optical response tensor according to the Kubo formula is given as follows:
\begin{equation}
  \begin{aligned}\label{eq:Kubotime}
      &\sigma^{\mu;\alpha_1,\alpha_2}(\omega_1,\omega_2)\\
      &=-\frac{e^3}{2\hbar^2} \int^0_{-\infty}dt_1 e^{-i(\omega_1+i\gamma)t_1} \int^{t_1}_{-\infty}dt_2 e^{-i(\omega_2+i\gamma)t_2}\\
      &\quad\,\times\mathrm{Tr}[\rho[r^{\alpha_2}(t_2),[r^{\alpha_1}(t_1),v^{\mu}]]]+((\omega_1,\alpha_1) \leftrightarrow (\omega_2,\alpha_2)),
  \end{aligned}
\end{equation}
where $\rho\,(\rho_{ab}=f_a \delta_{ab})$ is the density matrix and the square braket $[A, B]=AB-BA$ is the commutator.
Expanding the Bloch matrix representation of the commutation relations, the following relation is obtained~\cite{PhysRevB.99.045121,PhysRevX.11.011001}:
\begin{equation}
  \begin{aligned}\label{eq:ExcRelExp}
    [r^{\alpha_2}(t_2),[r^{\alpha_1}(t_1),v^{\mu}]]_{aa}&=i\partial_{\bm{k}}^{\alpha_2}i\partial_{\bm{k}}^{\alpha_1}iv_{aa}^{\mu}\\
    &\quad+i\partial_{\bm{k}}^{\alpha_2}[\xi^{\alpha_1}(t_1),v^{\mu}]_{aa}\\
    &\quad+[\xi^{\alpha_1}(t_1),(i\partial_{\bm{k}}^{\alpha_2}+t_1\Delta^{\alpha_1})v^{\mu}]_{aa}\\
    &\quad+[\xi^{\alpha_1}(t_2),[\xi^{\alpha_1}(t_1),v^{\mu}]]_{aa},
  \end{aligned}
\end{equation}
where
\begin{align}
  \Delta^{\alpha_1}_{ab}=\partial_{\bm{k}}^{\alpha_1}(\varepsilon_a-\varepsilon_b).
\end{align}
The integrals of the first, second, third, and fourth terms on the right-hand side of \Eq{eq:ExcRelExp} correspond to 
$\sigma_{\mathrm{ii}}, \sigma_{\mathrm{ei}}, \sigma_{\mathrm{ie}}$, and $\sigma_{\mathrm{ee}}$, 
respectively.

\bibliographystyle{apsrev4-2}
\bibliography{main.bbl}

\end{document}